\documentclass[dvips]{acta}
\usepackage{supertabular,lscape,epsfig}
\usepackage{amssymb}
\usepackage{amsmath}
\usepackage{multirow}
\usepackage{float}
\mathchardef\mhyphen="2D
\usepackage{placeins}
\DeclareSymbolFont{ppa}{OT1}{ppl}{m}{it}
\DeclareMathSymbol{\vv}{\mathalpha}{ppa}{'166}

\SetPages{257}{278}

\SetVol{75}{2025}

\newcommand{\doi}{3.4}

\begin{document}
\newcommand\pvalue{\mathop{p\mhyphen {\rm value}}}
\newcommand{\TabApp}[2]{\begin{center}\parbox[t]{#1}{\centerline{
  {\bf Appendix}}
  \vskip2mm
  \centerline{\small {\spaceskip 2pt plus 1pt minus 1pt T a b l e}
  \refstepcounter{table}\thetable}
  \vskip2mm
  \centerline{\footnotesize #2}}
  \vskip3mm
\end{center}}

\newcommand{\TabCapp}[2]{\begin{center}\parbox[t]{#1}{\centerline{
  \small {\spaceskip 2pt plus 1pt minus 1pt T a b l e}
  \refstepcounter{table}\thetable}
  \vskip2mm
  \centerline{\footnotesize #2}}
  \vskip3mm
\end{center}}

\newcommand{\TTabCap}[3]{\begin{center}\parbox[t]{#1}{\centerline{
  \small {\spaceskip 2pt plus 1pt minus 1pt T a b l e}
  \refstepcounter{table}\thetable}
  \vskip2mm
  \centerline{\footnotesize #2}
  \centerline{\footnotesize #3}}
  \vskip1mm
\end{center}}

\newcommand{\MakeTableH}[4]{\begin{table}[H]\TabCap{#2}{#3}
  \begin{center} \TableFont \begin{tabular}{#1} #4 
  \end{tabular}\end{center}\end{table}}

\newcommand{\MakeTableApp}[4]{\begin{table}[p]\TabApp{#2}{#3}
  \begin{center} \TableFont \begin{tabular}{#1} #4 
  \end{tabular}\end{center}\end{table}}

\newcommand{\MakeTableSepp}[4]{\begin{table}[p]\TabCapp{#2}{#3}
  \begin{center} \TableFont \begin{tabular}{#1} #4 
  \end{tabular}\end{center}\end{table}}

\newcommand{\MakeTableee}[4]{\begin{table}[htb]\TabCapp{#2}{#3}
  \begin{center} \TableFont \begin{tabular}{#1} #4
  \end{tabular}\end{center}\end{table}}

\newcommand{\MakeTablee}[5]{\begin{table}[htb]\TTabCap{#2}{#3}{#4}
  \begin{center} \TableFont \begin{tabular}{#1} #5 
  \end{tabular}\end{center}\end{table}}


\newcommand{\MakeTableHH}[4]{\begin{table}[H]\TabCapp{#2}{#3}
  \begin{center} \TableFont \begin{tabular}{#1} #4 
  \end{tabular}\end{center}\end{table}}

\newfont{\bb}{ptmbi8t at 12pt}
\newfont{\bbb}{cmbxti10}
\newfont{\bbbb}{cmbxti10 at 9pt}
\newcommand{\uprule}{\rule{0pt}{2.5ex}}
\newcommand{\douprule}{\rule[-2ex]{0pt}{4.5ex}}
\newcommand{\dorule}{\rule[-2ex]{0pt}{2ex}}

\begin{Titlepage}
\Title{Near-Contact Binaries on the Path to Contact Binaries}
\Author{K.~~S~t~ę~p~i~e~ń}{Astronomical Observatory, University of Warsaw, Al. Ujazdowskie 4,\\ 00-478 Warszawa, Poland}
\vspace*{-14pt}
\Received{November 11, 2025}
\end{Titlepage}
\vspace*{-14pt}

\Abstract{A comprehensive evolution study was conducted on a carefully
  selected sample of near-contact binaries (NCBs) with more massive
  components filling the Roche lobes, utilizing the best-known basic
  parameters and indications of ongoing mass transfer. The results and
  discussion highlight that several NCBs with total masses exceeding 2~\MS\
  survive only a short time after mass exchange as contact binaries (CBs),
  with both components eventually merging to form a rapidly rotating giant,
  akin to FK~Com. Less massive NCBs transition into typical CBs and
  remain in this phase for up to 2~Gyr before ending their binary evolution
  as systems with extremely low mass ratios, susceptible to Darwin
  instability.

  However, this does not fully explain the existence of low-mass CBs with
  masses in the range of 1-–1.5~\MS. It is noted that there exists a
  population of low-mass binaries, nearly filling their Roche lobes. Their
  overall properties suggest that they could be progenitors of low-mass
  CBs.}{binaries: close -- binaries: eclipsing -- Stars: evolution}

\vspace*{-9pt}
\Section{Introduction}
\vspace*{-5pt}
The current paper addresses the evolutionary state of near-contact binaries\break
(NCB) which appear to be related to contact binaries (CB), also known as
W~UMa-type binaries. CBs are believed to originate from solar-type detached
binaries in which angular momentum loss (AML) through magnetic braking (MB)
and stellar evolution effects lead to mass transfer. MB arises from
magnetically driven stellar winds (Huang 1966, Vilhu 1982, Stępień 1995),
which remove angular momentum and a small amount of mass, implying that CB
progenitors were initially somewhat more massive. Mass transfer begins when
the primary fills its Roche lobe while the secondary remains detached.

Many short-period, cool detached binaries and W~UMa-type stars are
observed, but the proposed evolutionary link requires intermediate
systems. The most promising candidates are binaries in which one or both
stars nearly fill their Roche lobes. Shaw (1990) termed these near-contact
binaries (NCBs) and divided them into two categories -- systems where the
primary fills its Roche lobe and those where the secondary does. Later,
Yakut and Eggleton (2005) proposed the designations SD1 and SD2 for the
first and the second category, respectively. SD2 binaries resemble\break

\newpage
\noindent
Algol-type systems, and indeed, no sharp boundary exists between
short-period Algols and SD2 stars.

This leads to the following evolutionary picture -- a detached binary loses
angular momentum {\it via} MB, shrinking the Roche lobe until the expanding
primary fills it when approaching the terminal-age main sequence
(TAMS). Roche lobe overflow (RLOF) initiates mass transfer, producing an
SD1 configuration. Continued transfer reverses the mass ratio (Case A
evolution), creating an SD2 system and possibly a brief contact
phase. Competing effects of mass transfer (orbit widening) and AML (orbit
shrinking) determine whether the system evolves into a CB or a short-period
Algol.

CB evolution continues under the influence of these processes until either
the components merge or an extreme mass-ratio system becomes Darwin
unstable. Early models proposed that CBs formed in contact on the zero-age
main sequence (ZAMS), but their observed properties -- nearly equal surface
brightness and Roche lobe filling -- contradict this idea for genuinely
young stars. This discrepancy, known as the Kuiper paradox (Kuiper 1941),
arises because the mass-radius relation for young stars does not match the
Roche geometry.

Lucy (1976) resolved this by introducing the thermal relaxation oscillation
(TRO) model, in which a CB remains in global thermal equilibrium while each
star oscillates around its Roche lobe size. The system alternates between
contact and broken-contact phases, with secular mass transfer gradually
decreasing the mass ratio. TROs were later extended to evolved binaries
(Yakut and Eggleton 2005), predicting a population of broken-contact
systems -- identified with NCBs or CBs in poor thermal contact (Lucy and
Wilson 1979). The model assumes that when a CB is in a contact phase, both
components conform exactly to the equipotential surface implying negligible
mass motions in the surface layers. This assumption has been challenged on
both theoretical and observational grounds (Stępień 2009, Rucinski
2025). NCBs may give us some clues about formation and stability of CBs.

Thus, the evolutionary status of NCBs remains uncertain -- are they
first-time systems evolving toward contact through Case A mass transfer, or
CBs temporarily detached in a TRO cycle? The present paper investigates the
early evolution of binaries leading to SD1-type NCBs, assuming they are
first-timers. Section~2 presents supporting evidence based on comparisons
between NCBs binaries and CBs. Section~3 analyzes period variations in the
selected SD1 systems, showing that observed secular timescales are longer
than the primaries' thermal timescales but close to those of the
secondaries -- consistent with slow, regulated mass transfer (Stępień and
Kiraga 2013). However, other mechanisms, including the influence of third
bodies, may also contribute to period changes. Section~4 derives the ZAMS
progenitors of these binaries, Section~5 outlines their subsequent
evolution, Section~6 discusses the results whereas the final section
presents the evolutionary relationship of NCB/CB systems and summarizes the
main conclusions.

\Section{Comparison of the Global Parameters of SD1 Binaries with CBs}
A detailed comparison of NCBs and CBs, based on accurate stellar parameters
for a number of variables of each type, was carried out by Yakut and
Eggleton (2005). They analyzed 72 CBs and 25 NCBs without distinguishing
between the SD1 and SD2 types. The authors noted ambiguity in the unique
assignment of several NCBs to either SD1 or SD2 type. Nevertheless, they
emphasized that a number of variables can be uniquely classified as SD1 or
SD2 and that both groups are comparably numerous.

They calculated the average total binary mass, mass ratio, orbital angular
momentum (AM), and luminosity of the primary component for both
groups. Based on these results, they concluded that all four quantities are
significantly lower for CBs than for NCBs. This finding apparently rules
out the possibility that NCBs, as a whole, are CBs in a broken-contact
phase of thermal oscillation. Instead, it suggests that NCBs are systems
experiencing mass transfer for the first time, following Roche lobe
overflow (RLOF) by the primary component -- a possibility that Yakut and
Eggleton (2005) acknowledge.

However, several SD1 binaries exhibit stellar parameters within the range
typical of CBs so could be related to them. On the other hand, the authors
noted that the rapid mass transfer following RLOF occurs on thermal
timescale of a primary star, which is much shorter than the time spent by
an SD2 binary in the Algol configuration after mass ratio
reversal. Consequently, we should observe far fewer SD1 than SD2 binaries,
which is contrary to observations. So Yakut and Eggleton (2005) came to the
conclusion that only a small fraction of SD1 binaries belongs to genuine
first-time mass transfer systems, while the rest must be broken-contact
oscillators. Yet they stressed that no clear criteria exist to distinguish
between these two cases.

A possible solution to the problem of similar numbers of both types of NCBs
was proposed by Stępień and Kiraga (2013) within the framework of a new
evolutionary model of CBs developed by Stępień (2006ab, 2009,
2011). According to this model, all CBs are formed as a result of mass
exchange in close binaries with mass ratio reversal -- \ie through Case A
binary evolution. SD1 binaries then represent the initial phase of mass
transfer, followed by a rapid merger of both components or the SD2 phase
leading to formation of either a CB or a short-period Algol, depending on
the amount of AM remaining in the system.

Using newer and more accurate data for CBs and NCBs, Stępień and Kiraga
(2013) compared not only the mean values of the global parameters but also
their distributions. They collected data on 22 SD1 and 27 SD2 binaries and
compared them with 110 CBs analyzed by Gazeas and Stępień (2008).

The comparison between NCBs of SD1- and SD2-types showed that the two
groups do not differ significantly from one another. The mean values of
total mass and orbital AM are equal within the limits of observational
accuracy, although SD2 binaries tend to have slightly higher angular
momentum. The only statistically significant difference was found in the
mean orbital periods -- $0.55\pm0.03$~d for SD1 and $0.67\pm0.03$~d for SD2
-- a result consistent with the somewhat higher AM of the latter group.

The close similarity between both groups is confirmed by the resemblance of
their mass and angular momentum (AM) distributions. For this reason,
Stępień and Kiraga (2013) combined the SD1 and SD2 binaries into a single
sample to compare them with CBs. This comparison revealed substantial
differences between the two groups.

The average total mass of NCBs was found to be $2.27\pm0.06$~\MS, while
that of CBs was $1.81\pm0.04$~\MS. The corresponding mean values of AM were
$(8.73\pm0.50) \times 10^{51}$ and $(4.68\pm0.25)\times10^{51}$ [cgs],
respectively, and the mean orbital periods were $0.61\pm0.02$~d for NCBs
and $0.42\pm0.01$~d for CBs. Moreover, the distributions of all three
parameters differ significantly. A $\chi^2$ test indicated that there is
only a negligible probability that the samples were drawn from the same
parent distributions.

The most recent compilations of data -- covering 48 NCBs (Meng \etal 2022)
and 437 CBs (Latković \etal 2021) -- confirm the previously reported
difference in total mass. The average total mass of NCBs is 2.29~\MS,
whereas for CBs it is only 1.56~\MS.  While the mean value for NCBs is very
close to that obtained by Stępień and Kiraga (2013), the value for CBs is
significantly lower (1.56~\MS\ \vs 1.81\MS). This discrepancy results from
a selection effect -- Gazeas and Stępień (2008) analyzed exclusively binaries
with spectroscopically determined masses, whereas most binaries compiled by
Latković \etal (2021) were characterized mostly
photometrically. Spectroscopic studies tend to target brighter -- and
therefore more massive -- systems.

We therefore conclude that NCBs represent a population distinct from that
of CBs. The differences in mass and orbital AM suggest that, if an
evolutionary connection exists between them, CBs must be the products of
NCB evolution. The lack of systematic differences between the SD1 and SD2
subtypes (apart from orbital period) indicates that both groups are in
closely related evolutionary phases whose durations are short compared to
the timescale of mass and AM loss.

\vspace*{-9pt}
\Section{Period Variations and Mass Transfer Rates of Selected SD1
  Variables}
\vspace*{-5pt}

In this section, we discuss the timescales of mass transfer between the
components of several SD1 variables and compare them with the
expected timescales inferred from the physical properties of each
binary. Throughout this analysis, we assume that the observed period
variations result solely from conservative mass transfer from the primary
to the secondary component. The purpose of this investigation is the
numerical confirmation of the suggestion by Stępień and Kiraga (2013)
that the mass transfer rate in NCBs of SD1-type is governed by the thermal
time scale of a secondary (less massive) component rather than a
primary. If so, it rules out SD1 variables being CBs in a broken-contact
phase.

We searched the most recent literature and selected a sample of 16 NCBs of
SD1-type with reliable component parameters and well-determined period
variations. The basic observational data for these systems are provided in
Table~1. Here $P$ denotes period, $\dot{P}$ rate of period change, $q$ mass
ratio equal to $M_2/M_1$ with $M, R$ and $L$ describing stellar mass,
radius and luminosity. The more massive primary component, which fills its
Roche lobe, is denoted with the subscript 1 and its less massive
companion with 2.
\renewcommand{\TableFont}{\scriptsize}
\MakeTableee{l@{\hspace{10pt}}c@{\hspace{6pt}}c@{\hspace{6pt}}c@{\hspace{10pt}}c@{\hspace{10pt}}c@{\hspace{9pt}}c@{\hspace{10pt}}c@{\hspace{10pt}}c@{\hspace{10pt}}c@{\hspace{10pt}}r@{\hspace{10pt}}c@{\hspace{7pt}}c}{12.5cm}{Basic observational parameters of the SD1 binaries} 
{\hline
\noalign{\vskip4pt}
star  & $P$ & $\dot{P}$            & $q$ & $T_1$ & $T_2$ & $M_1$ & $M_2$ & $R_1$ & $R_2$ & $L_1$ & $L_2$  & Ref.\\
      & [d] & $\times 10^{-7}~$d/y &     & [K]   & [K[   & [\MS] & [\MS] & [\RS] & [\RS] & [\LS] & [\LS]  & \\ 
\noalign{\vskip4pt}
\hline
\noalign{\vskip4pt}
V361 Lyr  & 0.31 &$-0.83$& 0.69 & 6200 & 4500 & 1.26 & 0.87 & 1.02 & 0.72 & 1.39 & 0.19  & 1 \\
V473 Cas  & 0.42 &$-0.76$& 0.49 & 5830 & 4378 & 1.00 & 0.48 & 1.19 & 0.83 & 1.47 & 0.23  & 2 \\
GR Tau    & 0.43 &$-0.42$& 0.22 & 7500 & 3434 & 1.45 & 0.32 & 1.49 & 0.71 & 6.33 & 0.06  & 2 \\
CN And    & 0.46 &$-1.40$& 0.39 & 6450 & 4726 & 1.43 & 0.55 & 1.48 & 0.95 & 3.42 & 0.41  & 2,3 \\
FT Lup    & 0.47 &$-1.77$& 0.47 & 6700 & 4651 & 1.87 & 0.82 & 1.64 & 1.13 & 4.88 & 0.54  & 3,4 \\
BS Vul    & 0.48 &$-0.24$& 0.34 & 7000 & 4632 & 1.52 & 0.52 & 1.54 & 0.93 & 5.13 & 0.36  & 2 \\
II Per    & 0.48 &$-0.75$& 0.38 & 5740 & 4464 & 0.95 & 0.36 & 1.31 & 0.84 & 1.68 & 0.25  & 5 \\
TT Cet    & 0.49 &$-0.50$& 0.43 & 7091 & 5414 & 1.57 & 0.68 & 1.55 & 1.04 & 5.47 & 0.84  & 2 \\
RT Scl    & 0.51 &$-1.29$& 0.43 & 7000 & 4820 & 1.63 & 0.71 & 1.59 & 1.01 & 5.47 & 0.50  & 6 \\
V878 Her  & 0.53 &$-1.85$& 0.44 & 6300 & 4243 & 1.55 & 0.69 & 1.62 & 1.12 & 3.72 & 0.37  & 7 \\
RU Eri    & 0.63 &$-0.34$& 0.54 & 6900 & 5106 & 1.37 & 0.73 & 1.73 & 1.27 & 6.11 & 0.99  & 9 \\
V1010 Oph & 0.66 &$-3.97$& 0.47 & 7500 & 5132 & 1.89 & 0.89 & 2.01 & 1.40 & 11.52& 1.22  & 2,3 \\
BL And    & 0.72 &$-0.24$& 0.38 & 7500 & 4830 & 1.80 & 0.70 & 2.13 & 1.35 & 12.93& 0.89  & 2 \\
IZ Mon    & 0.78 &$-2.06$& 0.39 & 8500 & 5120 & 2.01 & 0.78 & 2.48 & 1.49 & 28.93& 1.37  & 11 \\
V609 Aql  & 0.80 &$-0.78$& 0.70 & 6050 & 5000 & 1.05 & 0.74 & 1.84 & 1.47 & 4.09 & 1.22  & 8 \\
V388 Cyg  & 0.86 &$-4.11$& 0.37 & 8750 & 5543 & 2.08 & 0.79 & 2.52 & 1.54 & 33.54& 2.02  & 2 \\
\noalign{\vskip4pt}
\hline 
\noalign{\vskip4pt}
\multicolumn{13}{p{12.5cm}}
{References: (1)~Hilditch \etal (1997), (2)~Qian \etal (2020), (3)~Siwak \etal (2010), 
(4)~Lipari and Sisteró (1986), (5)~Zhu \etal (2009), (6)~Hilditch \etal (1986), 
(7)~Nelson \etal (2025), (8)~Tian and Chang (2020), (9)~Williamon \etal (2013),
(10)~Li \etal (2014), (11)~Yang \etal (2016).}
}

Because our goal is to examine the relationship between SD1 and W~UMa-type
variables, we included only binaries with total masses below 3 $M_\odot$
(more massive NCBs are very rare anyway). To ensure high-quality parameter
determination, we selected systems for which radial velocity curves have
been published or, in cases where only photometric observations are
available, systems with orbital inclinations exceeding 80$^o$.
Spectroscopic data are available for V361~Lyr, CN~And, FT~Lup, TT~Cet,
RT~Scl, V878~Her, V1010~Oph, and RU~Eri (the data for TT~Cet and RU~Eri are
incomplete). The systems V473~Cas, GR~Tau, BS~Vul, II~Peg, and TT~Cet
exhibit total eclipses.

It is crucial for our analysis to select binaries whose period variations
result from mass transfer between the components. Period variations in
close binaries can arise from several mechanisms, including the presence of
a third companion. It is well established that close binaries frequently
have distant tertiary companions (Tokovinin \etal 2006, Rucinski \etal
2007). Such companions can remove orbital AM from the inner binary
(Eggleton and Kiseleva-Eggleton 2006, Fabrycky and Tremaine 2007) and may
also induce apparent period variations through the light-time effect. When
observational data are limited, a segment of a sinusoidal signal in the
$O-C$ diagram can mimic a parabolic trend, potentially suggesting a linear
period change. To minimize this risk, we included only those binaries that
show a pronounced O'Connell effect in their light curves, assuming that
this asymmetry is caused by a mass-transfer stream impacting the surface of
the secondary component (Knote \etal 2022).

It should be noted that the value of $\dot{P}$ for V361 Lyr is larger than
that given by Hilditch \etal (1997), who analyzed the $O-C$ curve of this
system. They apparently made a mistake when calculating $\dot{P}$ from the
third term of the ephemeris -- their value should be divided by $P$ to
yield the correct $\dot{P}$, which is listed in Table~1 (see also Lister
2009).

The luminosities were calculated from the stellar radii and effective
temperatures, adopting a solar effective temperature of 5772~K.

The conservative mass transfer rate, $\dot{M_1}$, is related to the period
variation rate $\dot{P}$ by the relation:
$$\dot{M_1}=\frac{\dot{P}M_1M_2}{3P(M_1-M_2)}\,,\eqno(1)$$
and the resulting observed time scale of mass transfer is:
$$\tau_{\rm{obs}}=-\frac{M_1}{\dot{M_1}}\,.\eqno(2)$$
Here $\dot{M_1}$ is in units \MS/yr and $\tau_{\rm{obs}}$
in years (as all other time scales).

According to the TRO theory this time scale should be equal to the
Kelvin-Helmholtz (K-H) or thermal time scale of the primary component.

Regarding the K-H time scale there is some ambiguity about its value. Some
authors define it as the time needed to radiate away the thermal energy of
a star, $T$, assuming its present luminosity. Others use the total binding
energy $U$ instead. From virial theorem we have $T=-U/2$ so both values
differ by a factor of two. In addition, the binding energy of a star
depends on mass distribution within the star:
$$U=-\frac{\alpha{\rm G}M^2}{R}\,,\eqno(3)$$
where G is the gravitational constant and $\alpha$ depends on the density
distribution $\rho{(r)}$. For a polytropic sphere $\alpha=3/(5-n)$, where
$n$ is a polytropic index equal to 0 for a constant density sphere, 3/2 for
adiabatic sphere describing a convective star, and to 3 for an approximate
model of the Sun. Still larger values are for more strongly concentrated
objects. For our stars of interest, \ie solar-type and convective, the
coefficient $\alpha$ varies between 3/2 and 6/7. If we allow for additional
factor of 1/2 (using $T$ instead of $U$), we obtain a possible interval of
$\alpha$ between 3/2 and 3/7, which means a factor of 3.5 difference. Yet
the most commonly used value is 1. It will also be used here, but we should
keep in mind the described ambiguity.

So, we adopt the following expression for the K-H time scale:
$$\tau=\frac{{\rm G}M^2}{LR}=3.1\times10^7\frac{M^2}{RL},\eqno(4)$$
where solar units are used in the last term.

Mass transfer following RLOF by the primary component has been numerically
modeled by several authors, including Webbink (1976, 1977ab), Sarna and
Fedorova (1989), and Ge \etal (2010). The rate of mass transfer depends on
the response of both stellar radii to mass loss or gain and on the rate of
change of the critical Roche surface. All primary components listed in
Table~1 are sufficiently massive to possess only thin (if any) convective
zones. As a result, they shrink upon mass loss.

At the same time, their Roche lobes also contract due to the tightening of
the binary orbit caused by mass transfer from the more massive to the less
massive star. Detailed calculations show, however, that this shrinkage is
small for initial mass ratios exceeding 0.5, so the mass-transfer timescale
stabilizes at a value corresponding to the K-H timescale of the donor. For
lower initial mass ratios, the Roche lobe contracts rapidly, the
mass-transfer rate increases dramatically, and the process quickly reaches
the dynamical regime.

\renewcommand{\TableFont}{\footnotesize}
\renewcommand{\arraystretch}{1.1}
\MakeTableee{lc@{\hspace{17pt}}c@{\hspace{17pt}}c@{\hspace{0pt}}r}{12.5cm}{Comparison of the observed
  and the theoretical K-H timescales of primary components} 
{\hline
\noalign{\vskip4pt}
star & $\dot{M}$ & $\tau_{1,{\rm KH}}$ & $\tau_{\rm{obs}}$ & $\tau_{\rm{obs}}/\tau_{1,{\rm KH}}$ \\
     & [\MS/y]   & [y]                 & [y]               &  \\
\noalign{\vskip4pt}
\hline
\noalign{\vskip4pt}
V361 Lyr  & $-2.5\times 10^{-7}$ & $3.4\times 10^7$ & $5.0\times 10^6$ & 0.1 \\
V473 Cas  & $-5.6\times 10^{-8}$ & $1.7\times 10^7$ & $1.8\times 10^7$ & 1.1 \\
GR Tau    & $-1.3\times 10^{-8}$ & $6.7\times 10^6$ & $1.1\times 10^8$ & 17.0 \\
CN And    & $-9.1\times 10^{-8}$ & $1.2\times 10^7$ & $1.6\times 10^7$ & 1.3 \\
FT Lup    & $-1.8\times 10^{-7}$ & $1.3\times 10^7$ & $1.0\times 10^7$ & 0.8 \\
BSVul     & $-1.3\times 10^{-8}$ & $8.8\times 10^6$ & $1.2\times 10^8$ & 14.0 \\
II Per    & $-3.1\times 10^{-8}$ & $1.2\times 10^7$ & $3.1\times 10^7$ & 2.6 \\
TT Cet    & $-4.1\times 10^{-8}$ & $8.7\times 10^6$ & $3.8\times 10^7$ & 4.3 \\
RT Scl    & $-1.1\times 10^{-7}$ & $9.2\times 10^6$ & $1.5\times 10^7$ & 1.6 \\
V878 Her  & $-1.4\times 10^{-7}$ & $1.2\times 10^7$ & $1.1\times 10^7$ & 0.9 \\
RU Eri    & $-2.8\times 10^{-8}$ & $5.3\times 10^6$ & $4.9\times 10^7$ & 9.0 \\
V1010 Oph & $-3.4\times 10^{-7}$ & $4.6\times 10^6$ & $6.0\times 10^6$ & 1.2 \\
BL And    & $-1.3\times 10^{-8}$ & $3.5\times 10^6$ & $1.4\times 10^8$ & 33.0 \\
IZ Mon    & $-1.1\times 10^{-8}$ & $1.7\times 10^6$ & $1.8\times 10^7$ & 11.0 \\
V609 Aql  & $-8.1\times 10^{-8}$ & $4.4\times 10^6$ & $1.3\times 10^7$ & 2.9 \\
V388 Cyg  & $-2.0\times 10^{-7}$ & $1.5\times 10^6$ & $1.0\times 10^7$ & 6.7 \\
\noalign{\vskip4pt}
\hline
}
Based on the data in Table~1, we calculated mass-transfer rates and the
corresponding observed timescales using Eqs.(1--2), together with the K-H
timescales from Eq.(4). These values are listed in Table~2, while the last
column gives the ratio of the two.  With the notable exception of V361 Lyr
(discussed separately in Section~4.2), the observed timescales are at best
comparable to the stellar K-H timescales but are, in most cases, considerably
longer. This indicates that the observed mass-transfer rates are
substantially lower than theoretically expected. Moreover, they show no clear
dependence on the stellar parameters.

It is worth emphasizing that we consider only systems with well-determined,
finite values of $\dot{P}$. There exist NCBs showing no detectable period
variations, implying much lower -- if any -- mass-transfer rates. Perhaps the
most extreme example is CX Vir, whose orbital period has remained constant
for more than 90~yr (Kreiner \etal 2001), suggesting $\dot{P}<10^{-9}$~d
yr$^{-1}$. The resulting $\tau_{\rm{obs}}$ exceeds $10^9$~yr, giving the
ratio $\tau_{\rm{obs}}/\tau_{1,{\rm KH}}$ below 0.01. Notably, Siwak \etal
(2010) describe CX Vir as ``almost contact'', meaning that its secondary
component lies even closer to its Roche lobe than in typical NCBs.

We emphasize that the K-H timescales were computed from the currently
observed stellar parameters, whereas observations indicate that -- except
possibly for V361~Lyr -- mass transfer must have been ongoing long enough for
the secondary components (accretors) to have nearly filled their Roche lobes,
leading to the imminent formation of contact binaries. Ongoing mass loss from
the donor reduces not only its mass but also, quite substantially, its
apparent luminosity (Webbink 1976, Ge \etal 2010). This effect leads to an
apparent increase in the K-H timescale relative to its value at the onset of
RLOF.

Mass-transfer models predict that the accretor fills its Roche lobe after
roughly 0.1 $M_\odot$ of material has been transferred from the donor
(Webbink 1976, Sarna and Fedorova 1989). To estimate the stellar parameters
at the onset of RLOF, we therefore reversed this amount of mass -- moving
0.1~\MS\ from the accretor back to the donor -- while keeping the orbital
angular momentum constant. The observed parameters of V361 Lyr, particularly
the radius of the secondary, suggest that this system is still in the initial
phase of mass transfer, so assuming 0.1 $M_\odot$ already transferred likely
represents an upper limit.
\renewcommand{\arraystretch}{1.2}
\MakeTable{lccccrrc@{\hspace{0pt}}r}{12.5cm}{Basic parameters of the investigated
binaries at the Roche lobe over-flow by the primary component} 
{\hline
\noalign{\vskip4pt}
star & $M_{1,{\rm{RO}}}$ & $M_{2,{\rm{RO}}}$ & $P_{\rm{RO}}$ & $R_{1,{\rm{RO}}}$ & \multicolumn{1}{c}{age} & $L_{1,{\rm{RO}}}$ & $\tau_{1,\rm{RO}}$ & $\tau_{\rm{obs}}$/$\tau_{1,\rm{RO}}$\\
     & [\MS]             & [\MS]             & [d]           & [\RS]             & \multicolumn{1}{c}{[y]} & [\LS]             & [y] &  \\
\noalign{\vskip4pt}
\hline
\noalign{\vskip4pt}
V361 Lyr  & 1.36 & 0.77 & 0.36 & 1.16 & $1.07\times 10^9$ & 8.04  & $6.00\times 10^6$ & 0.8 \\
V473 Cas  & 1.10 & 0.38 & 0.64 & 1.68 & $6.67\times 10^9$ & 3.19  & $6.60\times 10^6$ & 2.7 \\
GR Tau    & 1.55 & 0.22 & 1.08 & 2.95 & $2.41\times 10^9$ & 12.75 & $1.90\times 10^6$ & 57.9\\
CN And    & 1.53 & 0.45 & 0.69 & 2.00 & $1.72\times 10^9$ & 8.60  & $4.00\times 10^6$ & 4.0 \\
FT Lup    & 1.97 & 0.72 & 0.59 & 1.93 & $5.5\times 10^8$  & 21.06 & $2.60\times 10^6$ & 3.9 \\
BSVul     & 1.62 & 0.42 & 0.75 & 2.20 & $1.50\times 10^9$ & 11.30 & $3.30\times 10^6$ & 36.4 \\
II Per    & 1.05 & 0.26 & 0.94 & 2.22 & $6.00\times 10^9$ & 1.84  & $5.00\times 10^6$ & 6.2 \\
TT Cet    & 1.67 & 0.58 & 0.66 & 1.97 & $1.13\times 10^9$ & 11.64 & $3.70\times 10^6$ & 10.3 \\
RT Scl    & 1.73 & 0.61 & 0.67 & 2.02 & $1.06\times 10^9$ & 13.30 & $3.40\times 10^6$ & 4.4 \\
V878 Her  & 1.65 & 0.59 & 0.70 & 2.04 & $1.28\times 10^9$ & 11.36 & $3.60\times 10^6$ & 3.1 \\
RU Eri    & 1.47 & 0.63 & 0.79 & 2.10 & $2.16\times 10^9$ & 7.69  & $4.00\times 10^6$ & 12.3 \\
V1010 Oph & 1.99 & 0.79 & 0.81 & 2.36 & $7.7\times 10^8$  & 25.60 & $2.00\times 10^6$ & 3.0 \\
BL And    & 1.90 & 0.60 & 0.97 & 2.69 & $1.05\times 10^9$ & 22.40 & $1.80\times 10^6$ & 77.8 \\
IZ Mon    & 2.11 & 0.68 & 1.02 & 2.87 & $8\times 10^8$    & 34.00 & $1.30\times 10^6$ & 13.8 \\
V609 Aql  & 1.15 & 0.64 & 0.94 & 2.11 & $6.07\times 10^9$ & 3.81  & $4.90\times 10^6$ & 2.7 \\
V388 Cyg  & 2.18 & 0.69 & 1.12 & 3.10 & $7.5\times 10^8$  & 40.00 & $1.40\times 10^6$ & 7.1 \\
\noalign{\vskip4pt}
\hline
}

The resulting component masses and orbital periods are listed Table~3. The
inferred orbital periods at the onset of RLOF are longer than the present
ones, ranging from 0.59~d to 1.12~d, again with the exception of
V361~Lyr. The donor radii -- assumed to be equal to their Roche lobe size
-- are also shown in Table~3. These are substantially larger than zero-age
main-sequence (ZAMS) radii, reflecting the evolutionary advancement of the
donor stars.

From the evolutionary tracks for solar-metallicity stars (Bressan \etal
2012), we derived age and luminosity of each donor at the onset of
RLOF. Using these parameters, we subsequently computed the K-H timescales of
the donors and compared them with the presently observed timescales from
Table~2. The results of the computations are listed in Table~3 with the
ratio of both timescales given in the last column. The K-H timescales at
RLOF are significantly shorter than those calculated from the present
parameters, thereby increasing the discrepancy between theoretical and
observed values.

We conclude that, for all systems except V361 Lyr, mass transfer proceeds at
rates about one to two orders of magnitude lower than those expected from the
donors' K-H timescales. V361 Lyr is a special case. Assuming that the
transferred mass lies between zero and 0.1~\MS, the observed timescale is
roughly comparable to, or somewhat shorter than, the K-H
timescale -- consistent with theoretical predictions for the initial phase of
mass transfer.

The K-H timescales of the accretors, computed from the presently observed
parameters, lie between $6\times10^6$~yr and $7\times10^7$~yr, several times
longer than those of the donors. V361 Lyr again stands out, with a value of
$1.6\times10^8$~yr. The corresponding K-H timescales of the accretors at RLOF
are $1-9\times10^8$~yr. This assumes that, due to their low masses, the
accretors remain near the ZAMS while the donors fill their Roche lobes.

\newpage
The observed mass-transfer timescales, ranging from $5.6\times 10^6$ and
$1.4\times10^8$~yr (see Table~1), generally fall between the currently
calculated K-H timescales of the accretors and those at RLOF, again except
for V361 Lyr.

In summary, the observed mass-transfer timescales in NCBs whose accretors
nearly fill their Roche lobes are longer, often much longer, than the K-H
timescales of the donors. This result cannot be reconciled with the TRO
theory. Instead, the observed timescales correspond more closely to the much
longer K-H timescales of the less massive accretors. In other words, the
observational data suggest that all these binaries are undergoing mass
transfer at a much slower rate than during the initial phase. This is
consistent with the suggestion of Stępień and Kiraga (2013) that the
mass transfer rate is related to the shrinkage rate of the accretor after new
mass has been accreted.

The data for V361 Lyr do not fit this pattern -- consistent with observations
showing that its accretor has not yet expanded in response to mass transfer
and remains deep in its Roche lobe -- indicating that the system is still in a
very early stage of mass exchange. This binary is therefore excluded from the
following discussion and treated separately as a special case.

\Section{Progenitors of the NCBs of SD1 Type}
The previous section demonstrated that the observed mass-transfer rates in
SD1-type NCBs are inconsistent with the predictions of the TRO theory, but
align well with the assumption that these binaries are first-time mass
transfer systems. In other words, they are currently undergoing Case A mass
transfer, which may subsequently lead to the formation of a short-period
Algol, a W~UMa-type contact binary, or even the immediate coalescence of the
components, depending on the amount of AM retained in the system after mass
ratio reversal (Stępień 2006ab, 2009, 2011).

If this is the case, we can now ask: What are the ZAMS progenitors of these
binaries? To estimate their ZAMS parameters, we applied the evolutionary
model of cool close binaries developed by the author to each system under
study.

\subsection{Essentials of the Evolutionary Model}
The evolution of a cool close binary is divided into three distinct phases.

In the first phase, an initially detached binary gradually tightens its orbit
through MB. At the same time, both components expand due to their intrinsic
stellar evolution, with the more massive primary doing so at a faster
rate. This phase ends when the primary fills its critical Roche lobe,
initiating RLOF and mass transfer to the companion.

The second phase encompasses the rapid mass-transfer stage, during which both
components are out of thermal equilibrium. The third phase begins once
thermal equilibrium is reestablished following mass ratio reversal. During
this stage, slow mass transfer continues, accompanied by AML, ultimately
leading either to the coalescence of both components or to their transition
into an Algol-type system.

The fundamental assumptions and governing equations of the model -- namely the
expression for the total binary AM, denoted $H_{\rm{tot}}$, Kepler's Third
Law, the approximate formula for Roche lobe radii, and two empirical
relations describing mass and AML due to MB -- are presented
and discusses in detail elsewhere (Stępień 2011, Stępień \etal 2017). Here,
we simply remind the last two for a better visualization of the whole
process:
$$\dot M_{1,2}=-10^{-11}R_{1,2}^2\,,\eqno(5)$$
$$\frac{\dd H_{\rm{tot}}}{\dd t}=\frac{-4.9\times10^{41}(R_1^2M_1 +R_2^2M_2)}{P}.\eqno(6)$$

The formulae are calibrated by the observational data on rotation of
single, magnetically active stars of various ages and empirically
determined mass-loss rates of single, solar-type stars. We assume that the
total mass and angular momentum loss of a binary system is a sum of the
losses from both components.

To describe the single-star evolution of each component during the first
and third phases, we employ {\sf PARSEC} evolutionary models with solar
metallicity ($Z=0.014$, Bressan \etal 2012).\footnote{\it
  https://stev.oapd.inaf.it/cgi-bin/cmd}

Both empirical loss equations are based on observations of late-type stars
possessing subphotospheric convection zones, where magnetic fields are
generated, giving rise to magnetic activity, hot coronae, and stellar
winds. Most primaries in the analyzed binaries are too massive to sustain
significant convection. However, the lower-mass secondaries are expected to
have strong magnetic fields and high levels of activity. In the lack of any
estimates of the influence of the magnetized star on its close companion we
simply assume that the cool secondaries induce magnetic fields in the
hotter primaries, leading to the formation of hot coronae and magnetized
winds in both components.  This assumption is supported by X-ray
observations from Shaw \etal (1996), who reported that the X-ray fluxes of
NCBs reach the saturation level (see also Szczygieł \etal 2008). To allow
for potentially lower magnetic activity in massive primaries, we
arbitrarily modified the loss formulae by assuming that the mass and AM
losses of stars with masses exceeding 1~$M_\odot$ cannot surpass those of a
solar-mass star.

To determine the initial parameters of the analyzed binaries, only the
first evolutionary phase needs to be computed in reverse -- from RLOF back to
ZAMS. Starting with the data provided in Table~3 (excluding V361~Lyr), we
iteratively add, at each negative time step, the amounts of mass and
angular momentum specified by the model. Simultaneously, we recalculate the
radius of the primary component, accounting for both its increased mass and
younger evolutionary age. The computation terminates when the star reaches
zero age (ZAMS).

Because of their small masses, the evolutionary changes in the radii of the
secondary components are negligible, and we therefore assume that their
current radii are equal to the ZAMS values. However, their masses are
adjusted according to the adopted mass-loss and mass-gain relations.

The resulting ZAMS parameters for all investigated binaries are listed in
Table~4.

\renewcommand{\arraystretch}{1}
\MakeTableee{lccccrcc}{12.5cm}{Initial (ZAMS) parameters of
  the progenitors of the investigated SD1 binaries}
{\hline
\noalign{\vskip4pt}
star & $M_1$ & $R_1$ & $M_2$ & $q_0$ & $H_0$ & $H_{RO}/H_0$ & $P_0$ \\
\noalign{\vskip4pt}
\hline
\noalign{\vskip4pt}
GR Tau    & 1.574 & 1.545 & 0.221 & 0.140 &  5.992 & 0.66 &  3.625\\
CN And    & 1.547 & 1.543 & 0.454 & 0.293 &  8.747 & 0.74 &  1.610\\
FT Lup    & 1.976 & 1.773 & 0.723 & 0.366 & 12.701 & 0.89 &  0.812\\
BS Vul    & 1.635 & 1.594 & 0.423 & 0.259 &  8.640 & 0.76 &  1.634\\
II Per    & 1.110 & 1.003 & 0.264 & 0.234 &  4.939 & 0.61 &  3.467\\
TT Cet    & 1.681 & 1.625 & 0.584 & 0.347 & 10.518 & 0.82 &  1.166\\
RT Scl    & 1.741 & 1.652 & 0.614 & 0.353 & 11.250 & 0.83 &  1.144\\
V878 Her  & 1.663 & 1.614 & 0.594 & 0.357 & 10.920 & 0.81 &  1.275\\
RU Eri    & 1.492 & 1.486 & 0.639 & 0.428 & 11.599 & 0.76 &  1.665\\
V1010 Oph & 1.998 & 1.764 & 0.795 & 0.398 & 15.485 & 0.89 &  1.102\\
BL And    & 1.910 & 1.709 & 0.604 & 0.316 & 13.012 & 0.86 &  1.481\\
IZ Mon    & 2.118 & 1.793 & 0.684 & 0.323 & 15.363 & 0.89 &  1.384\\
V609 Aql &  1.211 & 1.131 & 0.666 & 0.550 & 11.381 & 0.67 &  2.537\\
V388 Cyg &  2.188 & 1.814 & 0.694 & 0.317 & 16.249 & 0.91 &  1.472\\
\noalign{\vskip4pt}
\hline
}

The ZAMS progenitors exhibit several interesting properties. The average
orbital period is 1.83~d, which is in good agreement with the results of
Stępień (2011), who analyzed the observed period distribution of 421
detached cool close binaries with periods shorter than two days. The
distribution agreed well with the theoretical distribution obtained under
the assumption that the young cool binaries have a short-period limit of
2~d and as they age, their periods evolve due to AML. This
lower limit for the initial period arises from the fact that young close
binaries are not formed by the fission of a single protostar, but rather
through early fragmentation processes and/or Kozai cycles accompanied by
tidal friction (Boss 1993, Eggleton and Kisseleva-Eggleton 2006). The
present data confirm the existence of the short-period limit around 2~d.

The mean initial total mass is 2.22~\MS, which is typical for NCBs (see
Section 2), while the mean mass ratio, $q$, is relatively low at 0.33, with
only one system (V609 Aql) exceeding 0.5. These low initial $q$ values have
important implications for the subsequent evolution of the binaries (see
below).

\subsection{The Special Case of V361 Lyr}
The component radii of V361 Lyr do not conform to the evolutionary status
of the other NCBs. The primary, with a mass of 1.26~\MS, has a radius of
only $1.02~\RS$ (Hilditch \etal 1997), which is significantly smaller than
the ZAMS radius of a star of the same mass and solar composition. Only
metal-poor stars with $Z=0.001$ or less possess such small radii while
still on the ZAMS. The discrepancy is even more pronounced for the
secondary -- with a mass of 0.87~\MS, it has a radius of only 0.72~\RS,
whereas a metal-poor ZAMS star with $Z=0.001$ would have a radius of about
0.75~\RS. This raises the question of whether V361 Lyr could be a very
young, extremely metal-poor system or the calculated values are not
correct.

Recent data from Gaia DR3 give $T=6018$~K and ${\rm [Fe/H]}=-0.96$, while
LAMOST reports $T=4970$~K and ${\rm [Fe/H]}=-0.44$ (Qian \etal 2018). Both
datasets suggest a metallicity deficit, though not as severe as
required. The kinematical data do not indicate that the variable belongs to
the Galactic halo. Its height above the Galactic plane and its space
velocity both imply that V361 Lyr is a member of the Galactic disk
population. On the other hand, the recent observations by Gaia reveal the
presence of a close, optical component that is about 3~mag fainter than
V361~Lyr. Neglecting the third light when modeling the light curve leads to
incorrect stellar parameters. A new solution is therefore necessary.

Two further observational inconsistencies are worth noting. The period
change rate determined by Hilditch \etal (1997) was highly uncertain -- its
error exceeds 50\% of the reported value. Nevertheless, the data indicate a
shortening of the orbital period, consistent with the pronounced O'Connell
effect seen in the binary's light curve. The authors derived a
mass-transfer rate and claimed good agreement with the radiation excess
responsible for the O'Connell effect. However, their calculation of the
mass-transfer rate was incorrect (see Section 3), casting doubt on this
conclusion.

Moreover, the orbital period recently determined from Gaia DR3 photometry
(Gaia Collaboration 2022) is longer than that used by Hilditch \etal (1997)
Clearly, additional, high-precision observations and a thorough re-analysis
of all available data are urgently needed before a reliable model of this
unique binary can be established.

\vspace*{9pt}
\Section{Eventual Fate of the Investigated NCBs}
\vspace*{5pt}
To explore the possible future evolution of the discussed systems, their
complete evolutionary models were computed from the initial ZAMS stage
through the phase of mass exchange up to the end of the third evolutionary
phase. The evolution of the orbital period is crucial for determining the
final fate of a binary system.

Fig.~1 presents the behavior of the orbital period of each investigated NCB
throughout its entire lifetime. The black curves represent the period
evolution of eleven binaries listed in Table~4, excluding those
specifically named in the diagram. The colored curves correspond to the
named variables. Open circles denote the presently observed orbital period
values.
\begin{figure}[htb]
\centerline{\includegraphics[width=9.7cm]{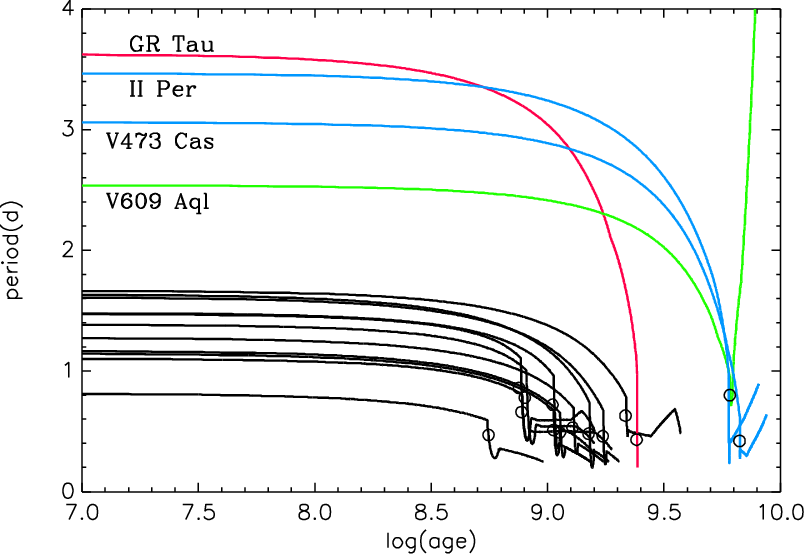}}
\vskip9pt
\FigCap{Orbital period evolution of investigated NCBs. Age [yr] from ZAMS
  is given on the abscissa. The colored curves, describing the behavior of
  the labelled binaries, are discussed individually in the text. Black
  curves describe all others stars from Table~4. Open circles mark the
  present values of the orbital periods.}
\end{figure}

All black curves display a broadly similar behavior. Due to the short
orbital period and low mass ratio, the orbit tightens substantially during
the phase of rapid mass transfer. After the mass equalization and a brief
SD2 phase, during which the period increases, the binary enters the contact
phase. In this stage, the mass transfer rate decreases while AML becomes
dominant, resulting in a shortening of the orbital period, overflow of the
outer critical Roche surface, and eventual merging of both components. In a
few cases, a brief interval of period increase is observed before the final
decrease. The calculations were terminated when both components
significantly overfilled the outer Roche lobe. The steep increase in mass
and AM loss during this stage leads to the coalescence of both components,
forming a rapidly rotating FK Com-type giant. The present model does not
adequately describe these late evolutionary stages.

To better illustrate the differences among individual systems, a portion of
Fig.~1 is enlarged and shown in Fig.~2. Note that the abscissa is
represented linearly here, in contrast to the logarithmic scale used in
Fig.~1. The segment of each curve between the SD2 phase and its end
represents the binary during its contact phase. As seen, the total binary
mass is the primary factor influencing the duration of this phase. This
relationship is demonstrated in Fig.~3, where the total presently observed
mass is plotted against the duration of the contact phase for all
investigated systems.
\begin{figure}[htb]
\centerline{\includegraphics[width=9.7cm]{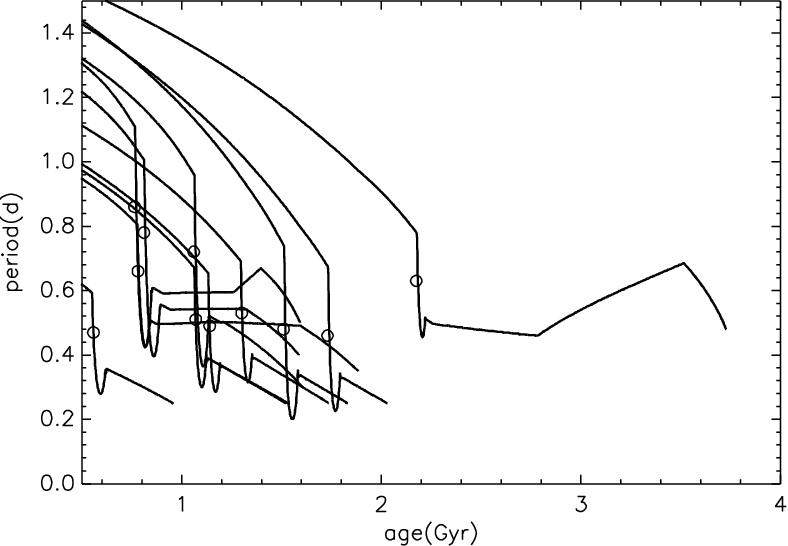}}
\vskip7pt
\FigCap{Zoomed in part of Fig.~1 with conversion of abscissa to
  linear scale to enhance differences among individual curves.}
\end{figure}
\begin{figure}[htb]
\centerline{\includegraphics[width=9.7cm]{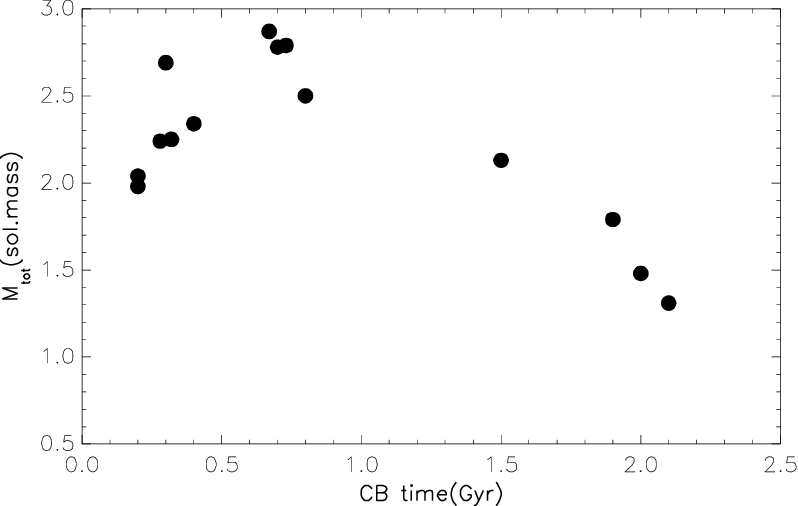}}
\vskip7pt
\FigCap{Presently observed total mass of the investigated binaries \vs
  duration of the contact phase of evolution until merging of both
  components. In case of V609~Aql it is until a transition to the Algol
  configuration with the orbital period of 5~d.}
\end{figure}

Two systems, II~Per and V473~Cas (blue curves in Fig.~1), exhibit a
different evolutionary outcome. The calculations indicate that their
components remain within the outer Roche lobe throughout the mass exchange
and the subsequent extended period of evolution. Slow mass transfer from
the former primaries dominates over AML, causing the orbital periods to
increase while the mass ratios decrease. The calculations were stopped when
$q$ reached 0.08, at which point the binary becomes susceptible to Darwin
instability, again leading to coalescence. Both systems spend approximately
2~Gyr as W~UMa-type binaries.

Finally, there is the case of V609 Aql (green curve). This system retains
sufficient AM to increase its orbital period during the third phase from a
minimum value of about 0.7~d up to the conventional upper limit for
W~UMa-type stars, thereby entering the domain of Algol-type binaries. The
calculations were terminated after 1.9~Gyr, when the period reached 5~d,
as the basic assumptions of the present evolutionary model apply only to
close binaries. The mass ratio of V609~Aql at this stage is 0.14. The
former primaries of all three systems develop small helium cores, each of
several hundredths of a solar mass, in the final phase. The corresponding
data for these three systems are plotted in Fig.~3 as the rightmost filled
circles.

\vspace*{4pt}

GR Tau does not fit the described evolutionary scenario. It has a very low
mass ratio. Assuming that it shares the same evolutionary history as the
other investigated binaries, its initial ZAMS mass ratio would have to be
even smaller -- only 0.14 (see Table~4). Such a binary would efficiently
lose AM already during the first (detached) phase, resulting in a
substantial orbital period shortening even before RLOF.

\vspace*{4pt}

Once mass transfer begins, the response of the fully convective
main-sequence secondary, with a mass of only 0.22~\MS, remains poorly
understood. Webbink (1977a) discussed the problem of mass accretion onto a
0.4~\MS companion with a very deep convective zone. Unfortunately, due to
computational difficulties, he was unable to follow the star's response
over sufficiently long timescales because of dramatic internal structural
changes and the formation of steep gradients in thermodynamic
parameters. Nevertheless, his calculations did not indicate any substantial
expansion. Later, Prialnik and Livio (1985) computed several models of mass
accretion onto a 0.2~\MS\ star under several simplifying assumptions and
concluded that a fully convective star can inflate substantially after
accreting less than 0.01~\MS, provided that the accretion rate is not
extremely low.

\vspace*{4pt}

GR Tau currently has a mass of 0.32~\MS and a radius of 0.71~\RS. If it is
indeed a first-timer and the presently observed mass transfer continues,
the orbital period will decrease to such a low value that overflow of the
outer Roche lobe will occur well before mass equalization, leading to the
merging of both components in less than $10^8$~yr. On the other hand,
the current parameters of the system are typical of a short-period
Algol. In particular, the evolutionarily advanced secondary with a mass of
0.32~\MS can easily reach the currently observed radius of 0.71~\RS.

\vspace*{4pt}

In fact, after analyzing the light curve, Lazaro \etal (1995) concluded
that GR Tau is a NCB of SD2-type with a transient O'Connell effect caused
by a cool stream of matter flowing from the low-mass component toward its
companion. A later analysis by Gu \etal (2004), however, indicated that GR
Tau is an SD1 binary with a decreasing orbital period. This is why it was
included in the present analysis, although it should be noted that many
W~UMa-type stars with poor thermal contact -- effectively non-contact
binaries (Siwak \etal 2010) -- can show either period decreases or
increases (Rucinski 2025).

Determining the precise evolutionary status and future fate of GR Tau
requires accurate spectroscopic observations and further detailed
investigation.

\section{Discussion of the Results}
It was demonstrated in Section 2 that the substantial differences between
the binary parameters of NCBs and CBs rule out the possibility that the
former are merely W~UMa-type stars in a semi-detached phase of the TRO
cycle. The higher values of orbital AM and total mass in NCBs instead
suggest that they may evolve into CBs after losing some mass and
AM. However, this raises the question of how they manage to lose sufficient
mass and AM before becoming W~UMa-type stars.

In the evolutionary sequence detached binary $\rightarrow$ SD1
$\rightarrow$ SD2 $\rightarrow$ CB, both the SD1 and SD2 stages are
short-lived compared to the total evolutionary lifetime of a cool close
binary. If a typical SD2 binary were to transform directly into a typical
CB, it would not have enough time to shed the necessary mass and AM. How,
then, can this problem be resolved?

A partial answer is provided by analyzing the final stages of NCB
evolution. As shown in the previous section, NCBs with initial total masses
higher than about 2~\MS follow the black curves in Fig.~1 and complete
their binary evolution along those tracks. Calculations indicate that, in a
fraction of these systems, soon after mass equalization during the mass
exchange phase, the outer critical Roche surface overflow occurs, resulting
in a prompt merger of both components.

The evolutionary model used in this paper does not allow for precise
computation of the remaining lifetime of these binaries from the present
time until coalescence. Nevertheless, the calculations suggest that the
overflow of the outer Roche lobe occurs when roughly half of the primary's
mass has been transferred, which typically takes a few
$\times10^8$~yr. After this stage, the systems form rapidly rotating giants
of the FK~Com type. Those with higher orbital AM survive longer -- up to
several $\times10^8$~yr -- as massive W~UMa-type variables with mass
ratios close to 0.4-0.5, but they also ultimately end their evolution as
mergers.

In contrast, the two binaries investigated with initial total masses lower
than 2~\MS -- evolving along the blue and green curves in Fig.~1 -- do not
overflow their outer Roche lobes until both components regain thermal
equilibrium. They then form typical W~UMa-type variables with orbital
periods shorter than 1~d, evolving toward extreme mass ratios, with
lifetimes on the order of $2\times10^9$~yr.

The last investigated binary, V609~Aql, possesses sufficient AM to increase
its orbital period beyond 1~d after mass exchange and becomes a
short-period Algol.

This, however, does not yet resolve the problem of the large number of
low-mass W~UMa-type binaries with total masses of about 1--1.5~\MS. We have
seen that they cannot originate from the known NCBs, as the latter systems
do not lose sufficient mass during either the SD1 or SD2 phases. A
population of low-mass NCBs is therefore needed to explain the origin of
these CBs.

A few such variables have been identified, including V1374 Tau, FS Aur, and
AD Cnc, with total masses of 0.99~\MS, 0.90~\MS, and 1.14~\MS,
respectively. However, the number of low-mass binaries classified as NCBs
remains far too small. It seems, though, that this is not a problem of
nonexistence but rather of accurately modeling observations to demonstrate
a variable's membership in the NCB class. Low-mass binaries are faint, so
obtaining high-quality photometry and spectroscopy is challenging.

In the section below we will discuss an evolutionary relationship of the
low-mass CBs with possible candidates for low-mass NCBs.

\Section{Evolutionary Relationship of the NCB/CB Systems}
Probable candidates for low-mass NCBs can be sought among binaries analyzed
by Pilecki (2010)\footnote{PhD Thesis, University of Warsaw}. He modeled
over 2000 carefully selected light curves of close eclipsing binaries
obtained within the ASAS project (Pojmański 2002). As a result, he derived,
among other parameters, the temperatures of the hotter and cooler
components ($T_h$ and $T_c$, respectively), their radii relative to the
sizes of the corresponding Roche lobes ($r_h$ and $r_c$), and approximate
mass ratios ($q$).

He classified the variables as CB, NCB, SD (semi-detached), or DB (detached
binary) based on the sum of the relative radii (rh + rc):

for $r_h+r_c>2$, the system was classified as CB,

for $1.9<r_h+r_c\le2$ as NCB,

for $1<r_h+r_c\le1.9$ as SD, and

for  $r_h+r_c\le1$ as DB.
\footnote{All data can be found in 
{\it http//www.astrouw.edu.pl/asas/?page=eclipsing}}

However, Pilecki (2010) recognized that his NCB classification was ambiguous,
so he always noted that a given binary could alternatively be classified as
CB and/or SD. It should be emphasized that the generally accepted definition
of an NCB requires that one component fills its Roche lobe, the other nearly
fills it, and that there be evidence of mass transfer between them. The
approximate models obtained by Pilecki (2010) do not, of course, fully
satisfy these conditions. Nevertheless, a substantial fraction of his NCBs
likely belong to the classical NCBs of the SD1 or SD2 type.

To identify low-mass NCBs among the stars investigated by Pilecki (2010) and
to compare them with other types of variables, all binaries classified as CB,
NCB, or DB with orbital periods shorter than 1~d were selected. To exclude
probable CBs from the sample of NCBs, an additional criterion was applied
based on the assumption that equal or nearly equal component temperatures
indicate very good thermal contact, characteristic of CBs. Therefore, all
binaries with temperature differences smaller than 10\% were rejected. The
10\% limit was chosen somewhat arbitrarily but rather conservatively, as
known NCBs typically show such or larger differences. However, a few
exceptional CBs are known to exhibit temperature differences of up to 10\%,
so it is possible that some CBs remain within the final NCB sample.

In total, 1,028 CBs, 165 NCBs, and 365 DBs were selected. The average
orbital periods are 0.53~d, 0.61~d, and 0.71~d for CBs, NCBs, and DBs,
respectively. The average values of $T_h$ are 6046~K, 6330~K, and 6128~K,
respectively, and the average mass ratios (defined as the ratio of the less
massive to the more massive component) are 0.395, 0.391, and 0.400. Fig.~4
shows the histograms of $T_h$ for all three groups of variables.
\begin{figure}[htb]
\centerline{\includegraphics[width=9cm]{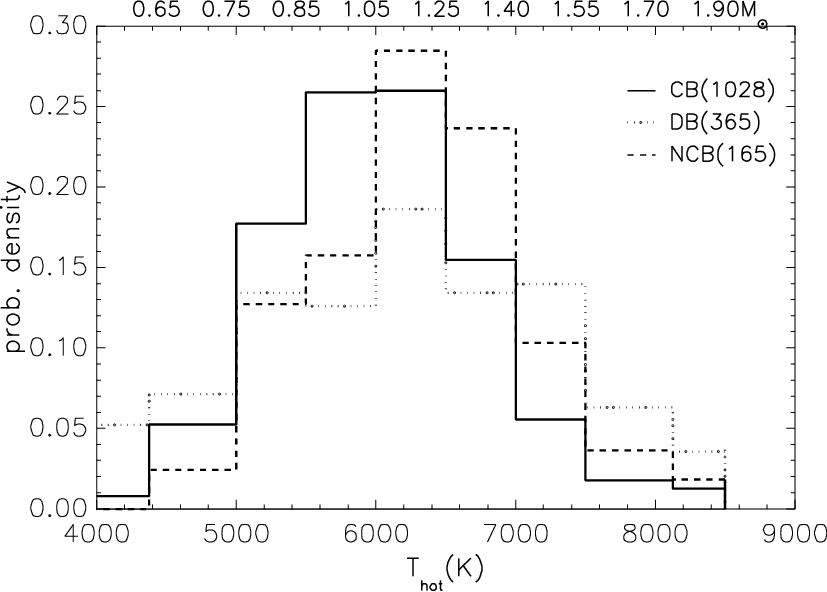}}
\vskip7pt
\FigCap{Histograms of the temperature of a hotter component normalized by
  the total number of variables of each class (indicated in
  parentheses) and coresponding masses of mainsequence stars.}
\end{figure}

Assuming that the hotter components are main-sequence stars, the
approximate masses corresponding to the indicated temperatures are also
shown. As can be seen, the temperature distributions are similar to each
other, indicating that the masses of the hotter components in all these
systems are also comparable. The nearly identical average values of $q$
suggest that there are no systematic differences in the masses of the
cooler components. Hence, the total masses are likewise similar.

Only the average orbital period increases systematically from CBs through
NCBs to DBs. This trend is confirmed by the period distributions of all
three groups of variables, as shown in Fig.~5. Although the differences are
not large (considering that only binaries with $P<1$~d were included),
they result in a systematic decrease of orbital AM from DBs to NCBs to
CBs. This is exactly what would be expected if the evolution of cool close
binaries were dominated by AML, leading to the formation of W~UMa-type
stars.
\begin{figure}[htb]
\centerline{\includegraphics[width=9cm]{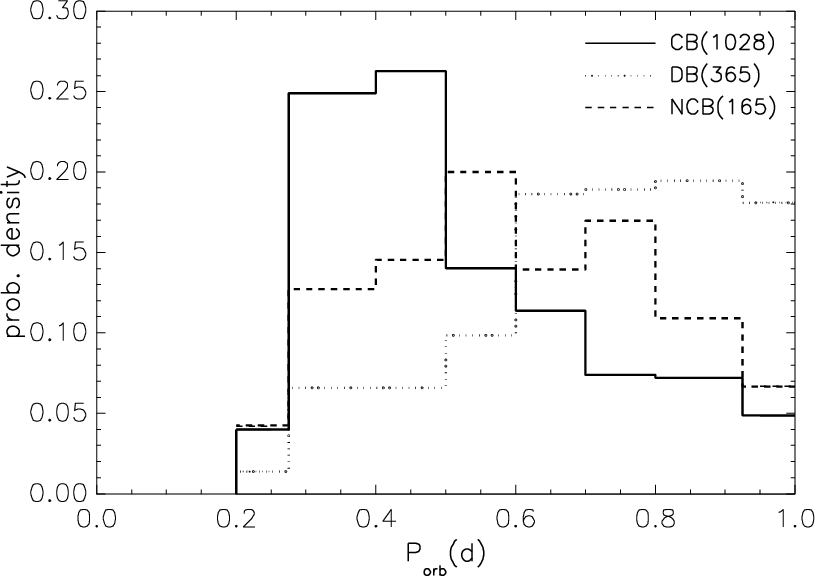}}
\vskip7pt
\FigCap{Histograms of the orbital period normalized by
  the total number of variables of each class (indicated in
  parentheses).}
\end{figure}

We can be more specific on this point. From Latković \etal (2021) we take
$M_1=1.18$~\MS\ and $M_2=0.38$~\MS\ as representative component masses of
CBs and assume identical masses for NCBs. With these values and above
derived periods we can calculate typical orbital AM of both groups. We obtain
$H_{\rm orb}({\rm CB})=3.88\times10^{51}$ and $H_{\rm orb}({\rm NCB})=
4.07\times10^{51}$. The AML of these binaries is equal to $\dot H_{\rm
  orb}= 0.02\times10^{51}/10^7$~yr, see Eq.(6). Consequently, the AM excess
of NCBs over CBs will disappear in about $9\times10^7$~yr, which is
remarkably close to $7\times10^7$~yr -- the estimated lifetime of the
binary modeled by Webbink (1976) in SD2 phase.
 
To sum up, the evolutionary computations indicate that the systematic
difference between known NCBs and CBs decreases when a large fraction of
massive NCBs is eliminated due to the early merging of their components. A
few massive NCBs with favorable initial parameters evolve into short-lived
A-type CBs. In contrast to massive NCBs, most lower-mass NCBs survive the
phase of rapid mass transfer and form long-lived CBs. In addition, a
population of low-mass NCB candidates has been identified among the close
binaries analyzed by Pilecki (2010), which may represent the main progenitors
of W-type CBs. High-precision spectroscopic observations are required to
accurately determine the evolutionary status of these variables.

The main uncertainty in the evolutionary computations of NCBs is related to
the adopted rate of angular momentum loss (AML). The formula used in this
study is based on the analysis of the spin-down of single, solar-type stars
possessing subphotospheric convection zones. Magnetic fields are generated
within these zones and interact with turbulent velocity fields, heating the
stellar coronae. The coronae then evaporate into interstellar space,
carrying magnetic fields with them. This process produces a drag force that
slows down the stellar rotation.

In the case of cool close binaries, it is assumed that the total AML rate
is the sum of the rates from both components. However, single main-sequence
stars with masses above 1.5~\MS\ do not possess subphotospheric convection
zones and therefore lack stellar winds capable of slowing their
rotation. It seems reasonable, nevertheless, to assume that a binary
component of such mass may possess a magnetic field induced by its less
massive companion and thus lose some of its own angular momentum.

The total initial mass is not the only parameter determining the fate of a
close binary. The initial mass ratio also plays a significant role. RU~Eri
has the highest mass ratio among all NCBs more massive than 2~\MS. As a
result, its orbital period does not shorten sufficiently during mass
transfer to cause the components to overflow the outer Roche lobe, and the
binary therefore remains a massive CB for about $3\times10^8$~yr before
merging. In contrast, GR Tau shows the opposite behavior -- although its
initial total mass is moderate (1.8~\MS), it has an extreme initial mass
ratio of only 0.14. Consequently, the orbital period decreases rapidly as a
result of mass transfer, leading to a prompt merger.

We conclude that not all binaries classified as NCBs will transform into
CBs. Only those with favorable component masses and orbital periods can
survive rapid mass transfer and remain in contact for an extended
period. Others merge, forming a rapidly rotating giant, or evolve into an
Algol-type binary. To identify a population of low-mass NCBs -- the
progenitors of W-type W~UMa binaries -- we need to observe low-mass,
short-period systems.

\Acknow{I would like to thank Slavek Rucinski for the interesting
  discussions, his careful reading of the manuscript, and the many remarks
  that significantly improved its presentation.

  This research has made use of the SIMBAD database operated at Centre de
  Données astronomiques de Strasbourg, France. This work also presents
  results from the European Space Agency (ESA) space mission Gaia. Gaia
  data are being processed by the Gaia Data Processing and Analysis
  Consortium (DPAC). Funding for the DPAC is provided by national
  institutions, in particular those participating in the Gaia MultiLateral
  Agreement (MLA). The Gaia mission website is
  {\it https://www.cosmos.esa.int/gaia} while the Gaia archive website is\newline
  {\it https://archives.esac.esa.int/gaia.}}

\end{document}